\documentclass[aps,prd,twocolumn, nofootinbib]{revtex4}

\usepackage{graphicx}% Include figure files
\usepackage{dcolumn}% Align table columns on decimal point
\usepackage{amssymb}% Outlined letters
\usepackage{hyperref}
\usepackage[all]{xy}
\usepackage[dvips]{epsfig}
\usepackage{amsfonts,amsmath}
\usepackage{caption}
\usepackage{subfig}
\DeclareGraphicsExtensions{.ps,.eps,.pdf,.png}

\usepackage[usenames,dvipsnames]{xcolor}

\newcommand{\half}{\frac{1}{2}}
\newcommand{\LCDM}{$\Lambda$CDM}
\newcommand{\al}{$\alpha$\ }
\newcommand{\omq}{$\Omega_{\rm q}$\ }

\begin{document}

\title{Cosmic microwave background constraints on coupled dark matter}
\author{Sophie C. F. Morris}
\email{ppxsm@nottingham.ac.uk}
\author{Anne M. Green}
\email{anne.green@nottingham.ac.uk}

\affiliation{School of Physics and Astronomy, 
University of Nottingham, Nottingham, NG7 2RD, UK} 

\begin{abstract}

We study CMB constraints on a scenario where a fraction of dark matter is non-minimally coupled to a massless scalar field and dark energy is in the form of a cosmological constant. In this case, there is an extra gravity-like fifth force which can affect the evolution of the Universe enough to have a discernible effect on measurements of cosmological parameters.  Using Planck and WMAP polarisation data, we find that up to half of the dark matter can be coupled. The coupling can also be several times larger than in models with a single species of cold dark matter coupled to a quintessence scalar field, as the scalar field does not play the role of dark energy and is therefore less constrained by the data. 
\end{abstract}

\maketitle

\section{Introduction}\label{sec:intro}

There is extensive observational evidence for the existence of dark matter, most of it favouring cold dark matter (CDM), see e.g. Ref.~\cite{Bertone:2004pz} for a review. In recent years there has been 
significant interest in models in which there are one or more species of DM coupled to a quintessence scalar field which plays the role of dark energy (e.g. Ref.~\cite{Farrar:2003uw,Brookfield:2007au,Bean:2008ac,Baldi:2012kt,Salvatelli:2013wra}).
Since the quintessence field must be light, this leads to additional long range forces on the DM, and the effects of such forces on structure formation and the cosmic microwave background (CMB) radiation have been studied~\cite{Friedman:1991dj,Gradwohl:1992ue,Frieman:1993fv,Gubser:2004uh,Gubser:2004du}. If these long range forces exist, but are neglected, they can affect measurements of the cosmological parameters~\cite{Kaloper:2009nc,2011PhRvD..84b3504D}.

It is possible that there is more than one species of DM (e.g. Ref.~\cite{Chialva:2012rq}).
Ref.~\cite{Brookfield:2007au} studied the phenomenology of multiple fluids interacting with different couplings to the dark energy scalar field, while Ref.~\cite{Baldi:2012kt} looked at the case of two species of dark matter, with identical physical properties, but different couplings to the dark energy. 
In Paper I~\cite{Morris:2013hua} we studied the cosmological evolution of a scenario with two species of DM, one of which is coupled to a scalar field. In contrast to, e.g., Ref.~\cite{Brookfield:2007au} we assume that the scalar field does not have a bare potential and the dark energy is in the form of a cosmological constant. This sort of set-up can, for instance, arise in higher dimensional compactifications~\cite{Garriga:1999yh}. 
This type of scenario can give rise to an abundance of scalar fields and, since the dark sector of our Universe is so poorly understood, there is plenty of scope for extra couplings to exist. It is therefore important to study the cosmological constraints on couplings in the dark sector, beyond just coupling of DM to quintessence
(in particular since quintessence may actually introduce more fine tuning problems than it solves, see Ref.~\cite{Copeland:2006wr} for a review). See Paper I for further details. We found that for the type of coupling studied in this paper, there is no longer a local minimum in the effective potential and there are no scalar field scaling solutions.
Consequently the evolution of the Universe can deviate substantially from that of $\Lambda$CDM. 
We found that a relatively small fraction of coupled dark matter can significantly modify the angular power spectrum of the CMB. In this paper we use the Planck temperature and WMAP polarisation data to constrain the fraction of coupled DM and its coupling strength. In Sec.~\ref{Desc} we briefly review the set-up. The CMB constraints are presented in Sec.~\ref{Fitting} and we conclude with discussion of the results in Sec.~\ref{Discussion}.

\section{Overview of scenario}
\label{Desc}

In this section we briefly overview the scenario which was introduced in Paper I~\cite{Morris:2013hua}. The matter content of the Lagrangian can be divided into two components, one of which consists of baryons, radiation and uncoupled dark matter, and the other which contains the remainder of the dark matter which is coupled to a massless scalar field, $\phi$. The action then takes the form

\begin{eqnarray}
\label{action}
S&=&\int {\rm d}^4 x \sqrt{-g}\left[\frac{1}{16\pi G}(R-2\Lambda)-\frac{1}{2}(\nabla\phi)^2\right] \nonumber \\ && +  \, S_{\rm SM}[g_{\mu\nu},...] + S_{\rm c}[g_{\mu\nu},...] + S_{q}[\hat g_{\mu\nu},\psi_{q}]\,,
\end{eqnarray}
where $g$ is the determinant of the metric $g_{\mu\nu}$, $R$ is the corresponding Ricci scalar and $\nabla$ is the covariant derivative. The action $S_{\rm SM}[g_{\mu\nu},...]$ contains the photons, baryons, and massless neutrinos and $S_{\rm c}[g_{\mu\nu},...]$ contains the uncoupled DM.
The ellipses in $S_{\rm SM}[g_{\mu\nu},...]$ denote the standard model fields, while the ellipses in $S_{\rm c}[g_{\mu\nu},...]$ denote the uncoupled dark matter field. Neither the standard model fields nor the uncoupled dark matter couple directly to the scalar.
The DM component that does couple directly to the scalar is described by the action $S_{q}[\hat g_{\mu\nu},\psi_{q}]$. The background evolution is governed by the Friedmann equation, 
\begin{equation}
H^2 = \frac{8\pi G}{3}\left(\rho_\Lambda+\rho_{\rm SM} +\rho_{\rm c} +\frac{1}{2}\dot\phi^2 + \rho_{\rm q} \right)\,,
\label{Friedmann}
\end{equation}
where $\rho_\Lambda$, $\rho_{\rm SM}$, $\rho_{\rm c}$ and $\rho_{\rm q}$
are the densities of the cosmological constant, the standard model fields (baryons, photons and neutrinos), uncoupled DM and coupled DM respectively and the scalar field equation of motion 
\begin{equation}
\ddot\phi+3H\dot\phi +\alpha\rho_q=0\,,
\label{KGeqn}
\end{equation}
where  $\alpha$ is the dimensionful coupling constant which determines the strength of the coupling between DM and the scalar field, 
 $H=\dot a/a$ is the Hubble parameter and overdots represent differentiation with respect to cosmic time $t$. Perturbing eqs.~\eqref{Friedmann} and \eqref{KGeqn} along with the background gravitational field gives 
\begin{align}
\delta'_q &= -\theta_q - \half h' + \alpha \delta\phi'  \,, \\
\theta'_q &= -\theta_q \mathcal{H} + \alpha(k^2\delta\phi - \phi'\theta_q) \, ,
\label{qcpert}
\end{align}
where $\delta\phi$ is the perturbation in the scalar field and ${\cal H}=a'/a$, where prime denotes differentiation with respect to conformal time. The perturbations of the scalar field are governed by 
\begin{equation}
\delta\phi'' + 2\mathcal{H}\delta\phi' + k^2\delta\phi +\half h'\phi' = -\alpha a^2 \delta \rho_q  \, .
\end{equation}

\section{Results}
\label{Fitting}

\begin{table}[t]
\centering
\begin{tabular}{|c|c|}
 \hline
 Parameter & Prior range \\
 \hline
 $\Omega_{\rm b} h^2$ & [0.005, 0.1] \\
  \hline
 $\Omega_{\rm c_{tot}} h^2$ & [0.001, 0.99]  \\
  \hline
 $100\theta_{*}$ & [0.5, 10.0]  \\
  \hline
 $\tau$ & [0.01, 0.8]  \\
  \hline
  $n_{\rm s}$ & [0.9, 1.1]  \\
  \hline
  $\ln(10^{10} A_{\rm s})$ & [2.7, 4.0] \\
    \hline
  $\Omega_{\rm q}$ & [0.0, 0.16] \\
    \hline
  $\alpha$ & [0.0, 0.3] \\
  \hline
\end{tabular}
\caption{Priors used for parameters in CosmoMC.}
\label{priors}
\end{table}

We use the Monte Carlo Markov Chain~\cite{Christensen:2001gj,Lewis:2002ah} code CosmoMC \cite{CosmoMC} to constrain 
the fraction of coupled DM and its coupling strength using the Planck and WMAP polarisation data. We use the nine year WMAP polarisation (WP)  data \cite{Bennett:2012zja} since it provides a tighter constraint on the optical depth than Planck is able to do using probes of gravitational lensing from large scale structure \cite{Ade:2013tyw}. Furthermore, the WMAP data is able to break some of the degeneracy between the matter density, $\Omega_{\rm m}= \Omega_{\rm b} + \Omega_{\rm c} + \Omega_{\rm q}$, and the Hubble constant~\cite{Ade:2013zuv}. We also use the same convergence criteria and sampling method as used in Ref.~\cite{Ade:2013tyw}.

In addition to the six \LCDM\ parameters (physical baryon density, $\Omega_{\rm b} h^2$, physical total cold dark matter density, $\Omega_{\rm c_{tot}} h^2= (\Omega_{\rm c} + \Omega_{\rm q}) h^2$, one hundred times the angular size of the sound horizon at decoupling, $100\theta_{*}$, optical depth at reionisation, $\tau$, scalar spectral index, $n_{\rm s}$ and primordial amplitude, $\ln(10^{10} A_{\rm s})$) we allow the density parameter of the coupled DM, $\Omega_{\rm q}$, and the coupling constant, $\alpha$, to vary~\footnote{It would also be possible to vary $\Omega_{\rm c} h^2$ and $\Omega_{\rm q}$ instead of $\Omega_{\rm c_{tot}} h^2$
and $\Omega_{\rm q}$, and this would give physically equivalent results. We chose the later pair of parameters in order to facilitate comparison with parameter constraints from $\Lambda$CDM, as most cosmological/astronomical probes are sensitive to the total CDM density.}. Our priors, all of which are flat, are given in Table~\ref{priors}. The density parameter of the coupled DM is allowed to vary in the range $0 < \Omega_{\rm q} < 0.16$ and the coupling in the range  $0 < \alpha < 0.3$. It is not possible to calculate the CMB angular power spectra for arbitrarily large values of $\Omega_{\rm q}$ and $\alpha$. As we saw in Ref.~\cite{Morris:2013hua}, there is no attractor solution for the scalar field. This means that small changes in the initial value of $\phi$ can lead to large changes in the present day densities, in particular for large $\alpha$. For large values of $\alpha$ it is not, in general, possible to find suitable initial conditions for the scalar field and furthermore, if the coupling is large the background cosmology changes too much unless the amount of coupled CDM is very small~\cite{Morris:2013hua}. Similarly for large \omq the evolution of the background deviates too much from $\Lambda$CDM to provide a good fit to the CMB data unless the coupling $\alpha$ is very small~\cite{Morris:2013hua}.  These regimes (where either $\Omega_{\rm q}$ or $\alpha$ are vanishingly small) are not physically interesting; it will never be possible to exclude a scenario in the limit where it tends to $\Lambda$CDM.

\begin{figure}[t]  
\includegraphics[width=0.4\textwidth]{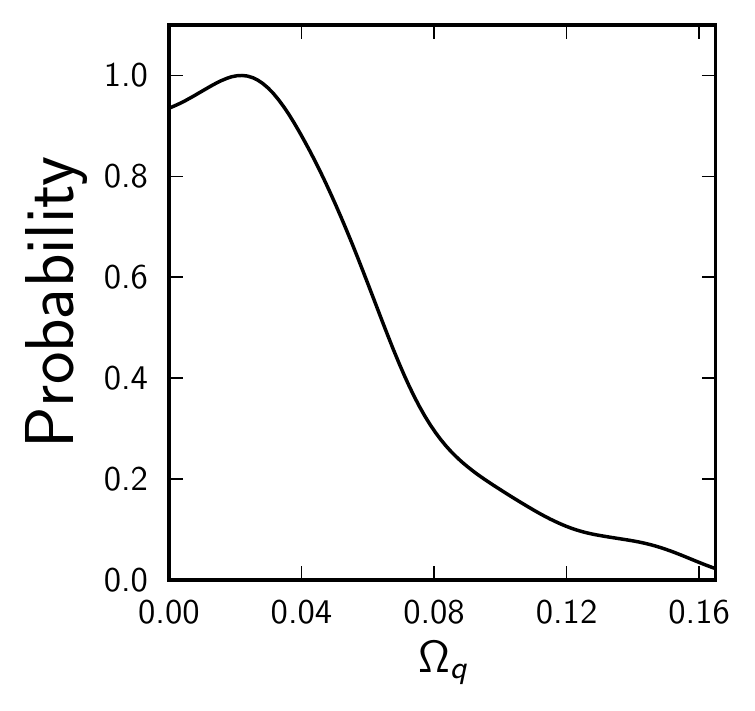} \\
\includegraphics[width=0.4\textwidth]{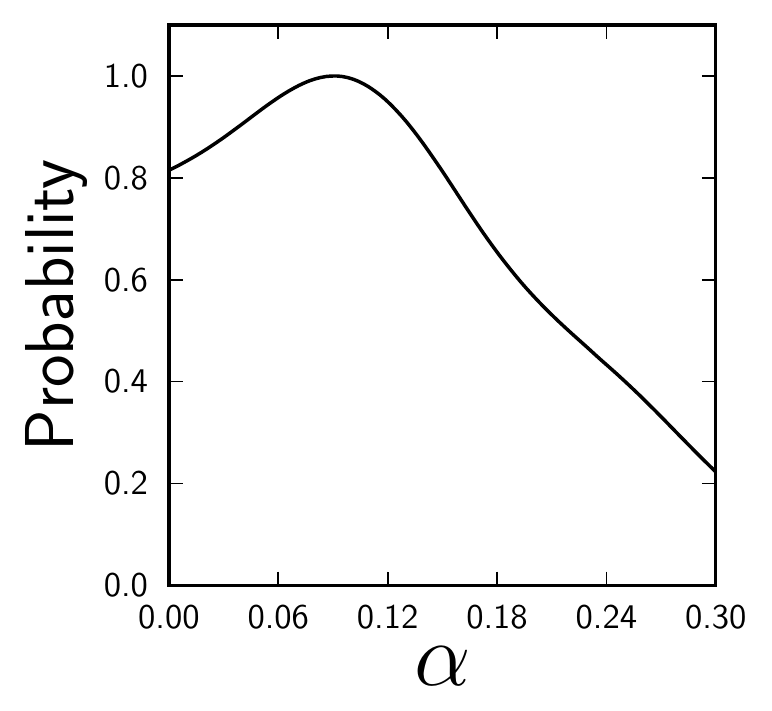}	
\caption{1D marginalised probability distributions for the coupled dark matter density,
 $\Omega_{\rm q}$, (top panel) and coupling constant, $\alpha$ (bottom).}
	\label{1Dplots}
\end{figure}

\begin{figure}[t!]
%\begin{center}
\includegraphics[width=0.5\textwidth]{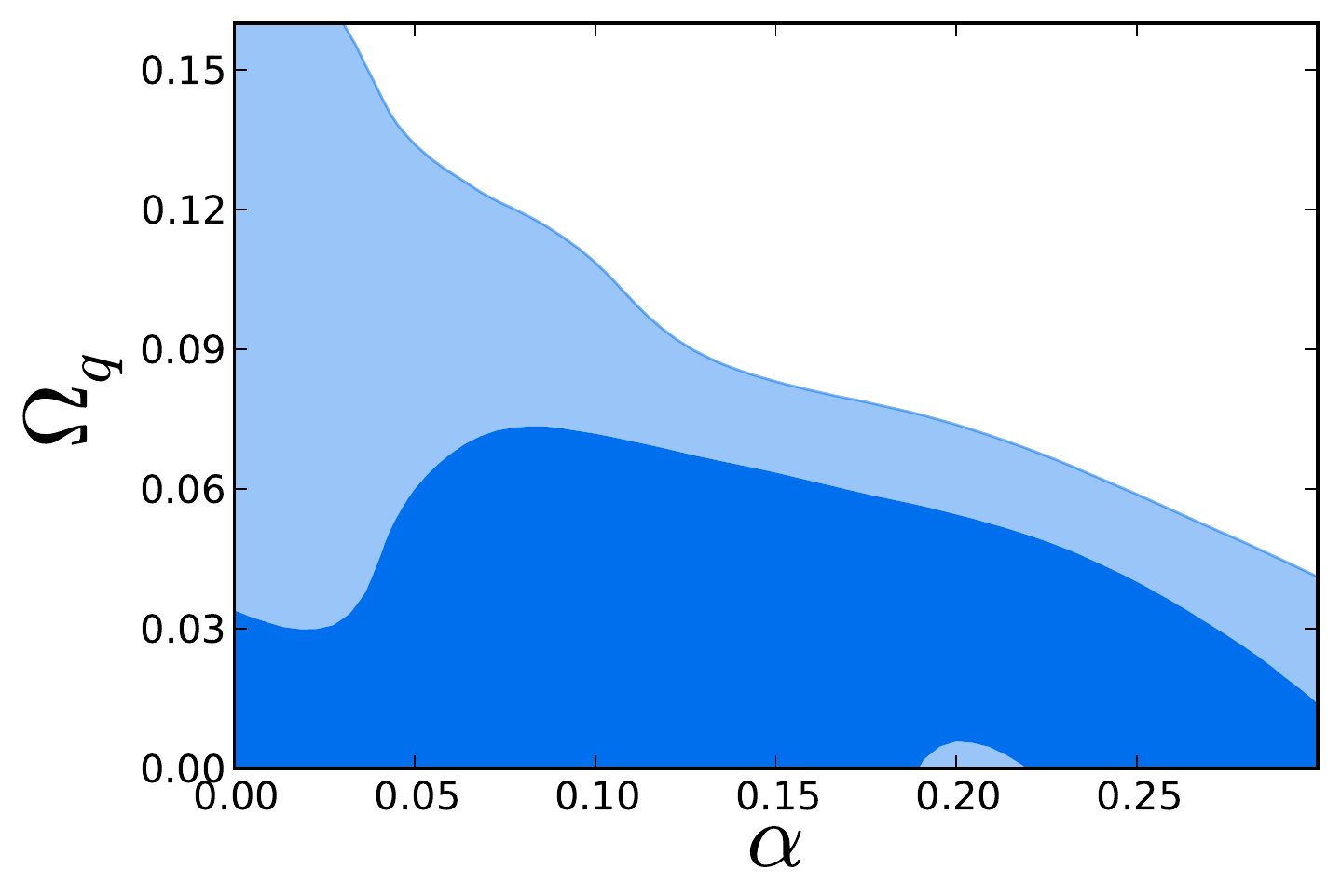}
\caption{Marginalized joint 68\% and 95\% confidence regions for \omq and $\alpha$. }
\label{degeneracy}
%\end{center}
\end{figure}

Fig.~\ref{1Dplots} shows the one-dimensional (1D)  marginalised probability distributions for the coupled dark matter density parameter, $\Omega_{\rm q}$, and coupling, $\alpha$, while Fig.~\ref{degeneracy} shows the  joint probability distribution of these parameters. The likelihood is maximised for non-zero values of  $\Omega_{\rm q}$ and $\alpha$, but there is no statistical preference for a non-zero coupled component. Large couplings and densities are, unsurprisingly, excluded and there is a degeneracy between \al and $\Omega_{\rm q}$; an increase in one of the parameters can be countered by a decrease in the other. However, the probability distributions of both parameters decrease fairly rapidly as their values increase. The fall off for $\Omega_{\rm q}$ is more rapid than that for $\alpha$. This is partly because, even for relatively small $\alpha$, if $\Omega_{\rm q}$ is large then the increase in the coupled DM density at early times leads to a large Integrated Sachs-Wolfe (ISW) effect at low multipoles~\cite{Morris:2013hua}. As discussed above, values of the coupling larger than $0.3$ would produce an acceptable fit to the CMB data, but only if the amount of coupled DM is small.

\begin{figure*}
%\begin{center}
\includegraphics[width=0.7\textwidth]{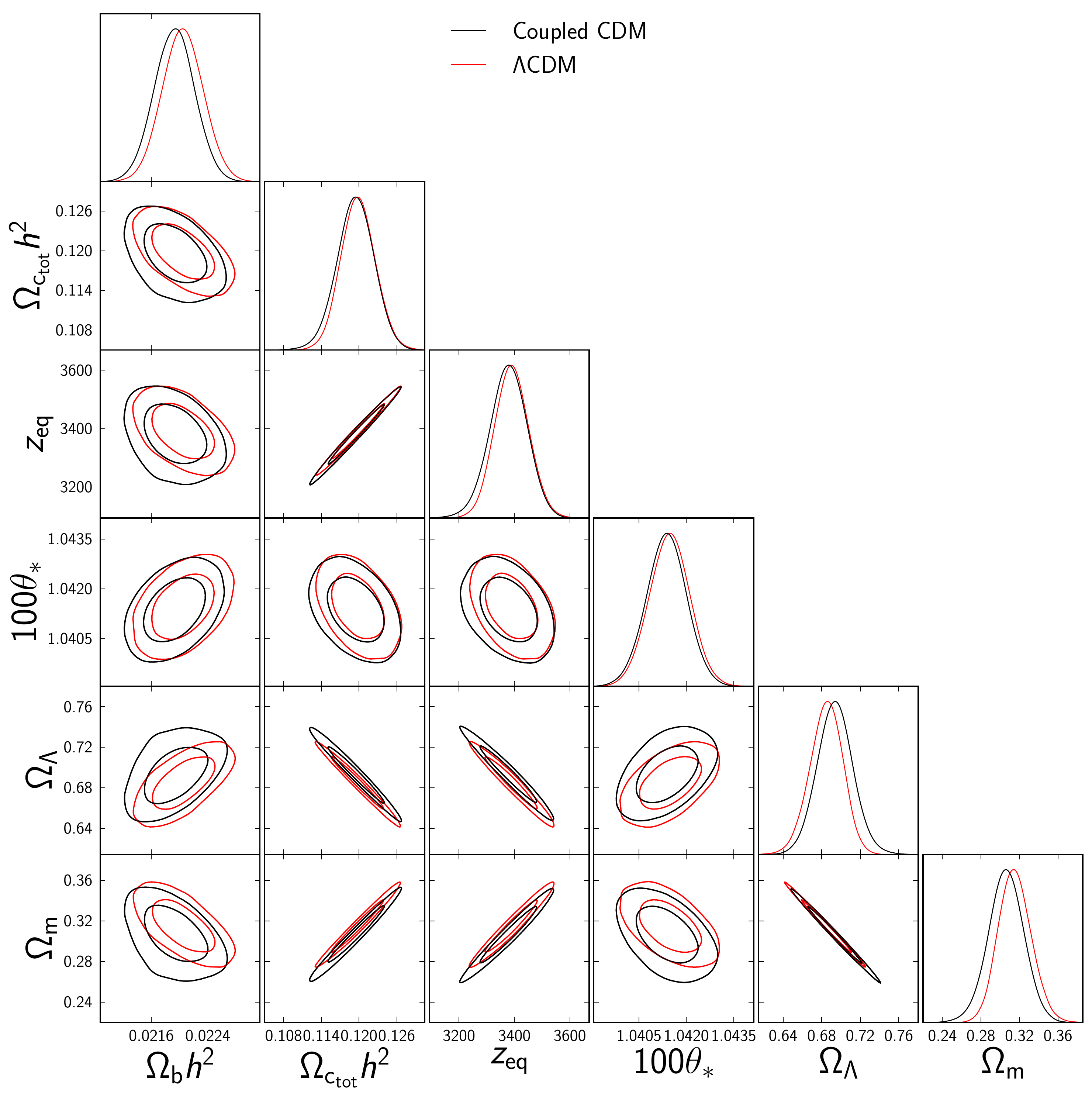}
\caption{1D marginalised probability distributions and corresponding 68\% and 95\% joint constraint contours for the total dark matter and baryon physical densities,   $\Omega_{\rm c_{tot}} h^2= (\Omega_{\rm q} + \Omega_{\rm c}) h^2$  and $\Omega_{\rm b} h^2$, the derived parameters $z_{\rm eq}$ and $\theta_{\star}$ and the matter and cosmological constant density parameters, $\Omega_{\rm m}$ and  $\Omega_{\Lambda}$,  from the Planck and WMAP polarisation data for coupled DM (dark black curve) and $\Lambda$CDM (light red curve).}
\label{Plancktriangle}
%\end{center}
\end{figure*}

 Fig.~\ref{Plancktriangle} compares the constraints on the density parameters for the coupled CDM scenario with those for \LCDM. It also shows the probability distributions of the derived parameters that have the most influence on the shape of the CMB angular power spectrum: the red-shift of matter radiation equality, $z_{\rm eq}$, and $\theta_*$. The constraints on $z_{\rm eq}$ are very tight and it is this, in addition to the ISW effect discussed above, which leads to the limits on the 
coupled DM density and coupling. As discussed in Paper I~\cite{Morris:2013hua}, if all other parameters are fixed, increasing the fraction of coupled dark matter increases the red-shift of equality.
The red-shift of equality can in principle be reduced by lowering the total matter density. However, this leads to an increase in the value of the scalar field and hence the density of the coupled DM. 
For modest coupled DM densities and couplings this can be compensated for by small shifts in the densities of the other components.  However, if the coupled DM density and coupling are too large then there is no combination of parameters which is compatible with the measured CMB peak heights and positions.

\section{Discussion}
\label{Discussion}

We have studied the constraints from Planck temperature and WMAP polarisation data on a scenario where there are two species of DM, one of which is coupled to a scalar field via a conformal coupling. 
There is a degeneracy between the density parameter of the coupled component and the coupling strength; a large coupling strength can be compensated for by a small density parameter and vice versa. From its ID marginalised probability distribution, the density parameter of the coupled DM is constrained to be $\Omega_q < 0.13$ at 95\% confidence. This corresponds to roughly half of the total CDM density. It is not possible to place an upper limit on the coupling, as a large coupling would be compatible with CMB data if the amount of coupled DM is vanishingly small. However, for significant coupled DM fractions the coupling can be of order 0.1, larger than is allowed for a single DM species coupled to a dark energy scalar field~\cite{Xia:2013nua,Li:2013bya}.

Other data sets may place tighter upper bounds on the fraction of coupled DM and its coupling strength. For example, the exchange of kinetic energy between the scalar field and coupled CDM could affect the abundance of DM halos \cite{Baldi:2010td}. Some work has been done using N-body simulations to simulate the non-linear regime of structure formation in models with multiple species of dark matter \cite{Baldi:2012ua}. These models with minimal coupling are essentially indistinguishable from \LCDM\ when looking at large-scale structure formation, but the separation of the different dark matter components starts to cause a fragmentation of structure for higher couplings. N-body simulations may, therefore, be useful in constraining the coupled DM fraction and its coupling strength, but are beyond the scope of this paper. Measurements of the Hubble parameter could also help to constrain this scenario. The forces between galaxies may change locally if the density of the coupled CDM varies from place to place, for example, an area containing many galaxies and an over density of coupled dark matter will feel an extra force coming from the CDM coupling. This will affect the recession velocity in this area, thus causing deviations between different measurements of the Hubble parameter. The potential magnitude of this effect has been explored in more detail in Ref.~\cite{Kaloper:2009nc}. Unfortunately the CMB data only places a constraint, in the context of coupled DM,  $(64.1 < H_0 < 71.8)\, \rm{kms^{-1} Mpc^{-1}}$ at  95\% confidence, and this is consistent with the locally measured value of $H_0 = 73.8 \pm 2.4\, \rm{kms^{-1} Mpc^{-1}}$ \cite{Riess:2011yx}.

\subsection*{Acknowledgments}

We would like to thank Adam Moss for useful discussions. 
SCFM and AMG are both supported by STFC.

%Hello World

\bibliographystyle{ieeetr}
\bibliography{bib_2_v5}

\end{document}